\documentclass[11pt,a4]{llncs} 

\usepackage[T1]{fontenc}
\usepackage[utf8]{inputenc}
\usepackage{amssymb,amsmath}
\usepackage[nonote]{marginote} 
\usepackage{times}
\usepackage{xspace}
\usepackage{enumitem}
\usepackage[disable]{todonotes}
\usepackage{microtype}

\newcommand\todoLR[1]{\todo[color=green!40]{\small #1}}
\newcommand\todoXU[1]{\todo[color=blue!40]{\small #1}}

\usepackage[left=3.4cm,top=4cm,bottom=4cm,right=3.4cm]{geometry}

\usepackage{url}
\usepackage{listings,subfigure}
\renewcommand{\ttdefault}{pcr}

\newcommand{\x}{\xspace}
\newcommand{\coq}{\textsc{Coq}\x}

\newcommand{\setN}{\ensuremath{\mathbb{N}}\x}
\newcommand{\setR}{\ensuremath{\mathbb{R}}\x}

\newcommand{\MAX}{\mbox{\textrm{max}}\x}
\newcommand{\SEC}{\mbox{\textsc{sec}}\x}

\newcommand{\DOM}{\mbox{\textrm{support}}\x}
\newcommand{\EQL}{\textrm{equilateral}\x}
\newcommand{\ISO}{\textrm{isosceles}\x}
\newcommand{\dest}{\textsl{dest}\x}
\newcommand{\target}{\textsl{target}\x}

\newcommand{\CEN}{\textrm{center}\x}
\newcommand{\BARY}{\textrm{barycenter}\x}
\newcommand{\IF}{\textbf{if}\x}
\newcommand{\THEN}{\textbf{then}\x}
\newcommand{\ELSE}{\textbf{else}\x}
\newcommand{\OR}{\textbf{ or }\x}
\newcommand{\BEGIN}{\textbf{begin}\x}
\newcommand{\END}{\textbf{end}\x}

\renewcommand{\epsilon}{\varepsilon}

\usepackage{lstcoq}
\usepackage{tikz}
\usepgflibrary{shapes.geometric}
\usetikzlibrary{shapes.geometric}
\usetikzlibrary{arrows.meta}

\tikzstyle{rond}=[circle=1pt,draw=black,line width=1pt]

\begin{document}



\title{Certified Universal Gathering in $\setR^2$ \\ for Oblivious Mobile Robots}

\institute{
  {\textsc{C\'edric} -- Conservatoire national des arts et
    m\'etiers, Paris, F-75141}
  \and {Collège de France, Paris, F-75006}
  \and{\'Ecole Nat. Sup. d'Informatique pour l'Industrie
    et l'Entreprise (ENSIIE), \'Evry, F-91025}
  \and {LRI, CNRS UMR 8623, Universit\'e Paris-Sud, Orsay, Université
    Paris-Saclay, F-91405}
  \and {UPMC Sorbonne Universit\'{e}s, LIP6-CNRS 7606}
\and Institut Universitaire de France
  }
\author{
Pierre Courtieu\inst{1}
\and Lionel Rieg\inst{2}
\and\\ Sébastien Tixeuil\inst{5,6}
\and Xavier Urbain\inst{3,4}
}

\maketitle

\begin{abstract}
  We present a unified formal framework for expressing mobile robots models, protocols, and proofs, and devise a protocol design/proof methodology dedicated to mobile robots that takes advantage of this formal framework.
  
  As a case study, we present the first formally certified protocol for oblivious mobile robots evolving in a two-dimensional Euclidean space. In more details, we provide a new algorithm for the problem of universal gathering mobile oblivious robots (that is, starting from any initial configuration that is not bivalent, using any number of robots, the robots reach in a finite number of steps the same position, not known beforehand) without relying on a common orientation nor chirality.
  We give very strong guaranties on the correctness of our algorithm by \emph{proving formally} that it is correct, using the \coq proof assistant.

  This result demonstrates both the effectiveness of the approach to obtain new algorithms that use as few assumptions as necessary, and its manageability since the amount of developed code remains human readable.  \bigskip
\end{abstract}

\section{Introduction}
\label{sec:introduction}

Networks of mobile robots captured the attention of the distributed computing community, as they promise new applications (rescue, exploration, surveillance) in potentially dangerous (and harmful) environments. Since its initial presentation~\cite{suzuki99siam}, this computing model has grown in popularity\footnote{The 2016 SIROCCO Prize for Innovation in Distributed Computing was awarded to Masafumi Yamashita for this line of work.} and many refinements have been proposed (see~\cite{flocchini12book} for a recent state of the art). From a theoretical point of view, the interest lies in characterising the exact conditions for solving a particular task.

In the model we consider, robots operate in Look-Compute-Move cycles.
In each cycle a robot ``Looks'' at its surroundings and obtains (in
its own coordinate system) a snapshot containing some information about the locations of all robots. Based on this visual information, the robot ``Computes'' a destination location (still in its own coordinate system) and then ``Moves'' towards the computed location. When the robots are oblivious, the computed destination in each cycle depends only on the snapshot obtained in the current cycle (and not on the past history of execution). The snapshots obtained by the robots are not necessarily consistently oriented in any manner.

The execution model significantly impacts the solvability of collaborative tasks. Three different levels of synchronisation have been considered. The strongest model~\cite{suzuki99siam} is the fully synchronised (FSYNC) model where each stage of each cycle is performed simultaneously by all robots. On the other hand, the asynchronous model~\cite{flocchini12book} (ASYNC) allows arbitrary delays between the Look, Compute and Move stages \todoLR{different for each robot} and the movement itself may take an arbitrary amount of time that is different for each robot. In this paper, we consider the semi-synchronous (SSYNC) model~\cite{suzuki99siam}, which lies somewhere between the two extreme models. In the SSYNC model, time is discretised into rounds and in each round an arbitrary subset of the robots are active. The active robots in a round perform exactly one atomic Look-Compute-Move cycle in that round. It is assumed that the scheduler (seen as an adversary) is fair in the sense that it guarantees that in any configuration, any robot is activated within a finite number of steps.

Designing and proving mobile robot protocols is notoriously difficult. The diversity of model variants makes it extremely onerous to check whether a particular property of a robot protocol holds in a particular setting. Even worse, checking whether a property that holds in a particular setting also holds in another setting that is not strictly contained in the first one often requires a completely new proof, even if the proof argument is very similar. The lack of proof reusability between model variants is a major problem for investigating the viability of new solutions or implementations of existing protocols (that are likely to execute in a more concrete execution model). Also, oblivious mobile robot protocols are mostly based on observing geometric constructions and deriving invariants from those observations. As the protocols are typically written in an informal high level language, assessing whether they conform to a particular model setting is particularly cumbersome, and may lead to hard to find mismatches. Hence, solely relying on handcrafted protocols, models and proofs is likely to introduce subtle errors that eventually lead to catastrophic failures when the system is actually deployed.
Formal methods encompass a long-lasting path of research that is meant to overcome errors of human origin. Not surprisingly, this mechanised approach to protocol correctness was successively used in the context of mobile robots~\cite{bonnet14wssr,devismes12sss,BMPTT13r,auger13sss,MPST14c,courtieu15ipl,berard15infsoc}.

\paragraph{Related Work.}

Model-checking proved useful to find bugs in existing literature~\cite{BMPTT13r} and assess formally published algorithms~\cite{devismes12sss,BMPTT13r}, in a simpler setting where robots evolve in a \emph{discrete space} where the number of possible positions is finite. Automatic program synthesis (for the problem of perpetual exclusive exploration in a ring-shaped discrete space) is due to Bonnet \emph{et al.}~\cite{bonnet14wssr}, and can be used to obtain automatically algorithms that are ``correct-by-design''. The approach was refined by Millet \emph{et al.}~\cite{MPST14c} for the problem of gathering in a discrete ring network. As all aforementioned approaches are designed for a discrete setting where both the number of positions and the number of robots are known, they cannot be used in the continuous space where robots positions take values in a set that is not enumerable, and they cannot permit to establish results that are valid for any number of robots. 

Developed for the \coq proof assistant,\footnote{\url{http://coq.inria.fr}} the Pactole framework enabled the use of high-order logic to certify impossibility results~\cite{auger13sss} for the problem of convergence: for any positive $\epsilon$, robots are required to reach locations that are at most $\epsilon$ apart. Another classical impossibility result that was certified using the Pactole framework is the impossibility of gathering starting from a bivalent configuration~\cite{courtieu15ipl}. While the proof assistant approach seems a sensible path for establishing certified results for mobile robots that evolve in a continuous space, until this paper there exists no \emph{positive} certified result in this context. Expressing mobile robot protocols in a formal framework that permits certification poses a double challenge: how to express the protocol (which can make use of complex geometric abstractions that must be properly defined within the framework), and how to write the proof?

\paragraph{Our contribution.}

Our first contribution is a unified formal framework for expressing mobile robots models, protocols, and proofs. This framework is motivated by the fact that many of the observed errors in published papers come from a mismatch between the advertised model and the model that is actually used for writing the proofs. For example, some dining philosophers protocols were expressed and proved in a high-level atomicity model, but advertised as working in a lower-level atomicity model, revealed to be incorrect in the lower-level atomicity model (see the work of Adamek \emph{et al.}~\cite{adamek15edcc} and references herein). Sometimes, the mismatch between the proof and the advertised model is more subtle: a perpetual exclusive exploration protocol the proof of which did not consider all possible behaviours in the advertised model ASYNC was used to exhibit a counter example in such a setting (See the work of Berard \emph{et al.}~\cite{BMPTT13r} and references therein). A unified formalisation whose consistency can be mechanically assessed is a huge asset for designing correct solutions, whose correctness can be certified. As we used a subset of the same framework for certifying impossibility results~\cite{auger13sss,courtieu15ipl}, consitency between negative and positive results is also guaranteed.
\todo{\small Insister sur le fait que c'est dans ce même framework qu'on a fait les preuves d'impossibilite? Ce qui ajoute
  un certain poids à l'affirmation d'universalité de notre algo.}

Our second contribution is a protocol design/proof methodology
dedicated to mobile robots. We advocate the joint development of both
the mobile robot protocol and its correctness proof, by taking
advantage of the \coq proof assistant features. The proof assistant is
typically able to check whether the proof of a particular
theorem/lemma/corollary is valid. So replacing particular clauses of
those theorems/lemmas/corollaries statements makes the proof assistant
check whether the proof still is acceptable for the new
statement. We used this feature to lift a preliminary version of this
paper (uni-dimensional setting~\cite{courtieu15corr}) to a Euclidean bi-dimensional space: the proof assistant checked which arguments 
were still
valid in the new setting. This feature also proved useful when
slightly changing parts of the algorithm: the impact of the changes on
the proofs were immediate. Also, it becomes easy to remove or weaken
hypotheses from the protocol, as the proof assistant makes it obvious if they are not used in the proof arguments.
Finally, our methodology includes a formal way to guarantee
whether the ``global'' view of the system (as seen from the protocol
prover point of view) is effectively realisable given the hypotheses
assumed in the model.

We instantiate our framework and methodology to actually design and
prove correct a new protocol for oblivious mobile robot universal
gathering problem. The mobile robot gathering problem is a
benchmarking problem in this context and can be informally defined as
follows: robots have \todoLR{j'ai enlevé « to coordinate »} to reach in a finite
number of steps the same location, not known beforehand. In more details, we
present a new gathering algorithm for robots operating in a continuous
space that \emph{(i)} can start from any configuration that is not
bivalent (that is, the robots are not initially equally placed in
exactly two locations, since gathering is impossible in this case),
\emph{(ii)} does not put restriction on the number of robots,
\emph{(iii)} does not assume that robots share a common chirality
(no common notion of ``left'' and
``right''). To our knowledge, this is the first certified positive
(and constructive) result in the context of oblivious mobile
robots. It demonstrates both the effectiveness of the approach to
obtain new algorithms that are truly generic facilitating the possibility to get rid of unnecessary assumptions,
and its manageability since the amount of developed code remains human readable. Our bottom-up approach permits to lay sound theoretical foundations for future developments in this domain.

\paragraph{Roadmap.}

Section~\ref{sec:formal} describes our formal framework, while our case study is developed in Section~\ref{sec:case}. Section~\ref{sec:discussion} gives some insights about the benefits of our methodology for mobile robot protocol design.

\section{A Formal Model to Prove Robot Protocols}\label{sec:formal}

To certify results and to guarantee the soundness of theorems, we use
\coq, a Curry-Howard-based interactive  proof assistant
enjoying a trustworthy kernel.
The (functional) language of \coq is a very expressive
$\lambda$-calculus: the \emph{Calculus of Inductive Constructions}
(CIC)~\cite{coquand90colog}. In this context, datatypes, objects, algorithms,
theorems and proofs can be expressed in a unified way, as terms.

The reader will find in~\cite{bertot04coqart} a very comprehensive overview
and good practices with reference to \coq.
Developing a proof in a proof assistant may nonetheless be tedious, or require
expertise from the user.
To make this task easier, we are actively developing (under the name
Pactole) a formal model, as
well as lemmas and theorems, to specify and certify results about
networks of autonomous mobile robots.
It is designed to be robust and flexible enough to express most of the
variety of assumptions in robots network, for example with reference
to the considered space: discrete or continuous, bounded or
unbounded\ldots

We do not expect the reader to be an expert in \coq but of course the
specification of a model for mobile robots in \coq requires some
knowledge of the proof assistant. %
We want to stress that the framework eases the developer's task. %
The notations and definitions we give hereafter should be simply read
as typed functional expressions. 

The Pactole model\footnote{Available at \url{http://pactole.lri.fr}} has been sketched
in~\cite{auger13sss,courtieu15ipl}; 
we recall here its main characteristics.

We use two important features of \coq: a formalism of
\emph{higher-order} logic to quantify over programs,
demons, etc., and the possibility to define \emph{inductive} and
\emph{coinductive} types~\cite{sangiorgi12book} to express
inductive and coinductive datatypes and properties.
Coinductive types are in particular of invaluable help to express
infinite behaviours, infinite datatypes and
properties on them, as we shall see with demons. 

Robots are anonymous, however we need to identify some of them in the
proofs. %
Thus, we consider given a finite set of \emph{identifiers}, isomorphic
to a segment of $\setN$. We hereafter omit this set \lstinline!G! 
unless it is necessary to characterise the number of
robots. 
Robots are distributed in space, at places called \emph{locations}.
We call a \emph{configuration} a \emph{function} from the set of
identifiers to the space of locations. %

From that definition, there is information about identifiers
contained in configurations, notably, 
equality between
configurations does \emph{not} 
boil down to the equality of the
multisets of inhabited locations. 

Now if we are under the assumption that robots are anonymous and
indistinguishable, we have to make sure that those identifiers are not
used by the embedded algorithm. %

\paragraph*{Spectrum.}\label{sec:spect}
The computation of any robot's target location is based on the
perception they get from their environment, that is, in an SSYNC
execution scheme, from a configuration. 
The result of this observation may be more or less accurate, depending
on sensors' capabilities. %
A robot's perception of a configuration is called a \emph{spectrum}. %
To allow for different assumptions to be studied, we leave abstract
the type \emph{spectrum} (\lstinline!Spect.t!) and the notion of
spectrum of a position. %
\emph{Robograms}, representing protocols, will then output a location when given a spectrum (instead
of a configuration), thus guaranteeing that assumptions over sensors
are fulfilled. %
For instance, the spectrum for anonymous robots with \emph{weak} global multiplicity detection could be the set of inhabited locations, i.e., without any multiplicity information.
In a \emph{strong} global multiplicity setting, the multiset of
inhabited locations is a suitable spectrum.

In the following we will distinguish a \emph{demon} configuration
(resp.  spectrum), 
expressed in the global frame of reference,
from a
\emph{robot} configuration (resp. spectrum), 
expressed in the
robot's own frame of reference. At each step of the distributed
protocol (see definition of \lstinline!round!  below) the demon
configuration and spectrum are transformed (recentered, rotated
and scaled) into the considered robots ones before being given as
parameters to the robogram. Depending on assumptions, zoom and
rotation factors may be constant or chosen by the demon at
each step, shared by all robots or not, etc.

\paragraph*{Demon.}\label{sec:demons}\todoXU{ nécessaire?}
Rounds in this SSYNC setting are characterised with set of oblivious robots
receiving their new frame of reference, if activated. We call
\emph{demonic action} this
operation together with the logical properties ensuring, for example,
that new frames of reference make sense.
\emph{Demons} are streams of demonic actions. As such, they are naturally
defined in \coq as a coinductive construct. Synchrony constraints
(e.g. fairness) may be
defined as coinductive properties on
demons, as detailed in~\cite{auger13sss,courtieu15ipl}.

The properties of fairness defined with inductive and
coinductive properties, can be understood rather
easily~\cite{auger13sss}.
\begin{lstlisting}
(** demon d activates at least once robot g. *)
Inductive LocallyFairForOne g (d : demon) : Prop :=
  | ImmediatelyFair : step (demon_head d) g <> None -> LocallyFairForOne g d
  | LaterFair : step (demon_head d) g = None -> 
                LocallyFairForOne g (demon_tail d) -> LocallyFairForOne g d.

(** A demon is Fair if at any time it will later activate any robot. *)
CoInductive Fair (d : demon) : Prop :=
  AlwaysFair : (forall g, LocallyFairForOne g d) -> Fair (demon_tail d) -> Fair d.

(** An execution is an infinite sequence of configurations*)
CoInductive execution : Type :=  
  NextExecution : Config.t -> execution -> execution
\end{lstlisting}

\paragraph*{Robogram.}\label{sec:robogram}
Robograms may be naturally defined in a \emph{completely abstract
  manner}, without any concrete code, in our \coq model.
They consist of an actual algorithm \lstinline!pgm! that represents the considered
  protocol and that takes a
spectrum as input and returns a location, and a compatibility property
\lstinline!pgm_compat! stating that target locations are the same if
equivalent spectra are given (for some equivalence on spectra).
\begin{lstlisting}
Record robogram :=
 {pgm :> Spect.t -> Location.t; 
  pgm_compat : Proper (Spect.eq \Parrow Location.eq) pgm}.
\end{lstlisting}

The generic notion of a round of the distributed algorithm considers
Byzantine robots (moved by the demon at each round). The robogram is
applied to each spectrum seen by each moving robot.
\begin{lstlisting}
Definition round (r: robogram) (da: demonic_action) (conf: Config.t) : Config.t :=
  fun id => (*@\hfill@*)(** for a given robot, we compute the new configuration *)
    match da.(step) id with (*@\hfill@*)(** first see whether the robot is activated *)
      | None => conf id (*@\hfill@*)(** If g is not activated, do nothing *)
      | Some sim => (*@\hfill@*)(** g is activated and [sim (conf g)] is its similarity *)
        match id with
        | Byz b => da.(relocate_byz) b (*@\hfill@*)(* Byzantine robots are relocated by the demon *)
        | Good g => (*@\hfill@*)(* configuration expressed in the frame of g *)
          let conf_seen_by_g := Config.map (sim (conf (Good g))) conf in
          (* apply r on spectrum + back to demon ref. *)
          (Sim.inverse (sim (conf (Good g)))) (r (Spect.from_config conf_seen_by_g))
        end
    end.
\end{lstlisting}

\section{Case study: A Universal Gathering for Mobile Oblivious Robots}
\label{sec:case}

The gathering problem is one of the benchmarking tasks in mobile robot networks, and has received a considerable amount of attention (see~\cite{flocchini12book} and references herein). The gathering tasks consists in all robots (considered as dimensionless points in a Euclidean space) reaching a single point, not known beforehand, in finite time.
A foundational result~\cite{suzuki99siam} shows that in the FSYNC or SSYNC models, no oblivious deterministic algorithm can solve gathering for two robots without additional assumptions~\cite{izumi12siam}. This result can be extended~\cite{courtieu15ipl} to the bivalent case, that is when an even number of robots is initially evenly split in exactly two locations. On the other hand, it is possible to solve gathering if $n>2$ robots start from initially distinct positions, provided robots are endowed with multiplicity detection: that is, a robot is able to determine the number of robots that occupy a given position.

While probabilistic solutions~\cite{suzuki99siam,izumi13tpds} can cope with arbitrary initial configuration (including bivalent ones), most of the deterministic ones in the literature~\cite{flocchini12book} assume robots always start from distinct locations (that is, the initial configuration contains no multiplicity points). Some recent work was devoted to relaxing this hypothesis in the deterministic case. Dieudonné and Petit~\cite{dieudonne12tcs} investigated the problem of gathering from \emph{any} configuration (that is, the initial configuration can contain arbitrary multiplicity points): assuming that the number of robots is odd (so, no initial bivalent configuration can exist), they provide a deterministic algorithm for gathering starting from any configuration. Bouzid \emph{et al.}~\cite{bouzid13icdcs} improved the result by also allowing an even number of robots to start from configurations that contain multiplicity points (albeit the initial bivalent configuration is still forbidden due to impossibility results in this case). In that sense, the algorithm of Bouzid \emph{et al.}~\cite{bouzid13icdcs} is \emph{universal} in the sense that it works for all gatherable configurations, including those with multiplicity points. The assumption that robots have a common chirality was removed in a context where robots may fail-stop in an unexpected manner~\cite{bramas15sirocco}.

A general description on how to characterise a solution to the problem
of gathering has been given in \cite{courtieu15ipl}. We specialise
this definition here to take into account that an initial
configuration is not bivalent. This is straightforward: any robogram
$r$ is a solution w.r.t. a demon $d$ if for
every configuration \emph{cf} that is not bivalent
(that is $\neg$ \lstinline!forbidden!), there is a point
$pt$ to which all robots will eventually gather (and stay) in the execution
defined by $r$ and $d$, and starting from \emph{cf}.

We present a new gathering algorithm for robots operating in a continuous space that \emph{(i)} can start from any configuration that is not bivalent, \emph{(ii)} does not put restriction on the number of robots, \emph{(iii)} does not assume that robots share a common chirality. We give very strong guarantees on the correctness of our algorithm by \emph{proving formally} that it is correct, using the \coq proof assistant. 
\begin{lstlisting}
Definition solGathering (r : robogram) (d : demon) :=
  forall cf, \not forbidden cf -> exists pt : R2, WillGather pt (execute r d cf).
\end{lstlisting}

\subsection{Setting and Protocol}\label{sec:unformal_robogram}

We consider a set of $nG$ anonymous robots that are oblivious and
equipped with global strong multiplicity detection (i.e., they are
able to count the number of robots that occupy any given position).
The demon is supposed to be fair, and the execution model is SSYNC.
The space in which robots move (the set of locations) is the real plane $\setR^2$; they do not share any common direction,
nor any chirality.
Any initial configuration is accepted as long as it is not bivalent (including those with multiplicity points).

\paragraph{Protocol.}\label{sec:unformal_algo}

The protocol we propose uses multiplicity to build the set of
towers of maximal height. %
If there is a unique tower of maximal height, i.e., a unique location
of highest multiplicity, this location is the destination of each
activated robot. %
Otherwise, the inhabited locations on the smallest enclosing circle (\SEC) are taken into account to define a \emph{target}. %

In our case robots enjoy strong global multiplicity detection: as noticed in Section~\ref{sec:spect} the spectrum of a
configuration is the multiset of all its robots' locations. %
It
is said to be \emph{clean} if inhabited locations are
either on the \SEC or at target. %
When 
it
is not clean (\emph{dirty}), robots on \SEC (or at
target) stay where they are, the others move to the target, thus
cleaning the spectrum. %
In a clean spectrum, any activated robot moves to the target. %
A configuration is said to be clean if and only if its spectrum is
clean. 

The important operation is thus to define a convenient target. %
Our target depends on how many inhabited locations are on the \SEC. %
If there is only one, then the whole spectrum 
is reduced to a single location and all robots are gathered. %
When the number of towers on the \SEC is not equal to 3,
the target is the center of the \SEC. 
Critical situations occur when towers on the \SEC define a triangle. %
If this triangle is \emph{equilateral}, the target is the center of
the \SEC (which is also the triangle's barycenter).
If it is \emph{isosceles} and not equilateral, the target is the
vertex opposite to its base. %
Finally if the triangle is \emph{scalene}, the target is the vertex opposite
to its longest side. %
Let us rephrase that description in informal pseudo-code. %
See Section~\ref{sec:robogram} for its formal version, that
is the \coq definition of our algorithm. %
For a spectrum $s$,
let $\DOM(s)$ be the set of locations in $s$, %
let $\MAX(s)$ be the set of locations of maximal multiplicity in $s$, %
and let $\SEC(s)$ be the smallest enclosing circle of $s$. %
Let \dest be the destination to be computed.
Remember that $(0,0)$ is always the location of a robot in its own frame
of reference.\medskip

{
\begin{tabular}{lll}
\multicolumn{3}{l}{\IF $\MAX(s) = \emptyset$ \THEN  $\dest:= (0,0)$ \hfill (* absurd case *)}\\
\ELSE & \multicolumn{2}{l}{\IF $\MAX(s) = \{p\}$ \THEN $\dest:= p$}\\
\ELSE & \multicolumn{2}{l}{\BEGIN \hfill (* first compute target then dest depending on
cleanliness *)}\\
     & \multicolumn{2}{l}{\IF $\DOM(s) \cap \SEC(s)  = \emptyset$ \THEN
       $\dest:= (0,0)$ \hfill (* absurd case *)}\\
     & \ELSE \IF &$\DOM(s) \cap \SEC(s) = \{p\}$ \THEN $\target:= p$ \hfill (* already gathered *)\\
     & \ELSE \IF &$\DOM(s) \cap \SEC(s) = \{p_1,p_2,p_3\}$ \THEN
     \hfill (* triangle cases *)\\
     &        & \IF $\EQL(p_1,p_2,p_3)$ \THEN $\target:= \BARY(p_1,p_2,p_3)$\\
     &        & \ELSE \IF $\ISO(p_1,p_2,p_3)$ \THEN $\target:= \textrm{opposite of base} (p_1,p_2,p_3)$\\
     &        & \ELSE  $\target:= \textrm{opposite of longest} (p_1,p_2,p_3)$\\

     & \multicolumn{2}{l}{\ELSE $\target:= \CEN(\SEC(s))$;}\\
     & \multicolumn{2}{l}{\IF $\forall p \in s, p \in \SEC(s) \OR p =
       \target\ $ \THEN $\dest:=\target$ \qquad (* clean $\Rightarrow$ go to target *)}\\
     & \ELSE \IF & $(0,0) \in \SEC(s) \OR (0,0) = \target\ $  
                  \THEN  $\dest:= (0,0)$ \hfill (* dirty $\Rightarrow$ clean config *)\\
     &  \multicolumn{2}{l}{\phantom{else} \ELSE $\dest:=\target$}\\
     & \END\\
\end{tabular}
}

\paragraph{Phases of the algorithm.}
We characterise several cases of the protocol, called \emph{phases}, which depend on what is perceived from the configuration, and which are mutually
exclusive in an execution: Gathered robots, the
Majority case where there is a unique tower of maximal height, the
three triangle cases (Equilateral, Isosceles, Scalene), and finally
the General case. %
To ease the proof of termination, we chose to consider differently an instance
of the general case, namely the Diameter case where
$\DOM(s) \cap \SEC(s)$ contains exactly two points (in which case they are a diameter of the \SEC).

\begin{figure}[htbp]
\centering
    \begin{tikzpicture}[thick,>={Stealth[length=3.5pt]},outer sep = 1pt]

      \node (n) at (12,-0.4) [color=black,shape=ellipse,ultra thick,draw] {Gathered}; %
      \node (n0) at (12,-1.85) [shape=circle,draw] {Maj}; %
      \node (n1) at (10.1,-1.2) [color=black,shape=circle,draw] {Dc}; %
      \node (n2) at (8.6,-1.2) [color=black,shape=circle,draw] {Dd}; %
      \node (n3s) at (7,-0.4) [shape=circle,draw] {Sc}; %
      \node (n3i) at (7,0.4) [color=black,shape=circle,draw] {Ic}; %
      \node (n3e) at (7,-1.2) [shape=circle,draw] {Ec}; %
      \node (n4s) at (5.5,-0.4) [color=black,shape=circle,draw] {Sd}; %
      \node (n4i) at (5.5,0.6) [shape=circle,draw] {Id}; %
      \node (n4e) at (5.5,-1.45) [shape=circle,draw] {Ed}; %
      \node (n5) at (2.5,-0.4) [shape=circle,draw] {Gc}; %
      \node (n6) at (1,-0.4) [shape=circle,draw] {Gd}; %

      \draw [color=black,line width=.2pt,->] (n0) to (n); %
      \draw [line width=.2pt,->] (n1) to (n); %
      \draw [color=black,line width=.2pt,->] (n2) to (n1); %
      \draw [color=black,line width=.2pt,->] (n3e) to (n2); %
      \draw [color=black,line width=.2pt,->,out=25,in=-200] (n3e) to (n); %
      \draw [color=black,line width=.2pt,->] (n4e) to (n3e); %
      \draw [color=black,line width=.2pt,->] (n4s) to (n3s); %
      \draw [color=black,line width=.2pt,->] (n4i) to (n3i); %
      \draw [color=black,line width=.2pt,->,out=5,in=140] (n3i) to (n); %
      \draw [color=black,line width=.2pt,->,out=15, in=-210] (n3s) to (n); %
      \draw [color=black,line width=.2pt,->,out=-15,in=-190] (n5) to (n3e); %
      \draw [color=black,line width=.2pt,->,out=15,in=-170] (n5) to (n3i); %
      \draw [color=black,line width=.2pt,->,out=-40,in=-130] (n5) to (n2); %
      \draw [color=black,line width=.2pt,->,out=-45,in=-130] (n5) to (n1); %
      \draw [color=black,line width=.2pt,->,out=35, in=200] (n5) to (n4i); %
      \draw [color=black,line width=.2pt,->,out=0,in=180] (n5) to (n4s); %
      \draw [color=black,line width=.2pt,->] (n6) to (n5); %

      \draw [color=black!50,->,>=stealth,out=20,in=140,dashed,very thick] (6.2,1) to (n0); %
      \draw [color=black,line width=.2pt,->,out=-40, in=-155] (n6) to (n0); %
      \draw [color=black,line width=.2pt,->,out=-30, in=-180] (n2) to (n0); %
      \draw [color=black,line width=.2pt,->,out=-55, in=-162] (n5) to (n0); %
      \draw [color=black,line width=.2pt,->] (n1) to (n0); %

      \draw [color=black!50,dashed,very thick] (4.95,1) rectangle (7.6,-1.9);
    \end{tikzpicture}
  \caption{Reachability graph for the distinguished categories of spectra.
    For clarity's sake, self loops are omitted.
    The boxed area contains the triangle cases; they are all linked to Maj.}
\label{fig:graphe}
  \end{figure}
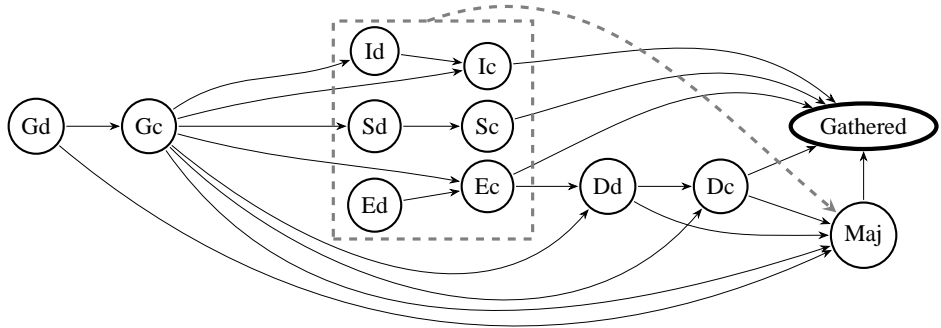

For all cases that need the computation of a target, we
moreover distinguish between clean and dirty situations.  Note that
from any dirty version of a case, the only two other reachable cases
are its clean version and Majority.  This leaves us with twelve phases:
Gathered (the success situation), Majority (Maj), Diameter clean (Dc)
and dirty (Dd), Equilateral, Isosceles, Scalene clean (Ec, Ic, Sc) and
dirty (Ed, Id, Sd), and General clean (Gc) and dirty (Gd).
Figure~\ref{fig:graphe} summarises the reachability relation between
cases. 

\subsection{Key points to prove correctness}
\label{sec:key-points}

Some properties are fundamental in our proof that the algorithm
actually solves Gathering. Namely, that robots move towards
the same location, that a legal configuration cannot evolve into a
forbidden (that is: bivalent) one, and finally that the configuration is
eventually reduced to a single inhabited location.

\paragraph{Expressing the robogram in the global frame of
  reference.}\label{sec:globalbehavior} %
The first step towards reasoning about a robogram is to leave the robots local frames of reference and rephrase the robogram in the demon global frame of reference.
This step is always left implicit in pen-and-paper proofs but it is actually not trivial:
it relies on the fact that the protocol uses only geometrical concepts that are invariant under the allowed changes of frame, here scaling, rotation, and translation.
Using a formal framework ensures that this overlooked proof is indeed
done and correct. %
This in turn gives a global version of the \lstinline!round! function and creates a global view of the configuration after one round (lemma \lstinline!round_simplify!).

\paragraph{Robots that move go to the same location.}\label{key:same_dest}
Note that by robots ``that move'' we explicitly mean robots that change location
during the round, \emph{not} robots that are activated (some of which may not move).
Robots enjoy global strong multiplicity detection, hence they all detect the number of highest towers, they share the same notion of \SEC, and they all compute the same number of towers on the \SEC.
Moreover, in both non-equilateral triangle cases, pointing out the longest side or the base side is not ambiguous as relative distances compare the same way for all robots.
Hence, cleanliness and targets are the same for all activated robots, which means that computed destinations are the same.

Further note that we actually just showed that all moving robots are
in the same phase of the protocol, %
and that the resulting destination does not depend on the frame of
reference of the robot. %

\paragraph{Bivalent positions are unreachable.}\label{key:not_biv}
We require that the initial configuration does not consist of exactly
two towers with the same multiplicity.  One of the key points ensuring
this algorithm's correctness is that there is no way to reach a position
that is bivalent from a position that is \emph{not}
bivalent. 
 Consider two configurations $C_0$ and $C_1$, $C_1$ being bivalent and resulting from
 $C_0$ by some round. Let us denote by $|x|_0$ (resp. $|x|_1$) the
 multiplicity of location $x$ in $C_0$ (resp. in $C_1$).
 By definition, $C_1$ consists of two locations $l_1$ and $l_2$ such that $|l_1|_1 =
 |l_2|_1=\frac{nG}{2}$. As all moving robots go to \emph{the same location}, we can assume
 without loss of generality that robots moved to, say, $l_1$, adding to
 its original multiplicity $|l_1|_0$ (which might have been $0$). 
 Since the configuration is now bivalent, this means that 
 $l_2$ was inhabited in $C_0$  and such that $|l_2|_0 \geq \frac{nG}{2}$ (some robot in $l_2$ might have
 moved to $l_1$). There cannot have been only one inhabited location $l$ distinct
 from $l_2$ before the round because either $|l|_0=|l_2|_0=\frac{nG}{2}$ but we supposed
 the configuration was not bivalent, or $|l|_0< \frac{nG}{2} < |l_2|_0$ but then by
 phase~Majority robots would have moved to $l_2$ and not $l_1$.
 Hence $C_0$ consisted of $l_2$ and several inhabited $l_i$ ($i\neq2$) amongst which the
 robots not located in $l_2$ were distributed, but then none of the
 $l_i$ could have held more than $\frac{nG}{2} -1$ robots, hence
 phase~Majority should have applied and robots should have moved
 to $l_2$, a contradiction.
Interestingly, this argument 
makes no reference to the dimension of the space.
It applies as is to both~\cite{courtieu15corr} and this work.

\paragraph{Eventually no-one moves.}\label{key:termination}
The termination of the algorithm is ensured by the existence of a
measure decreasing at each round involving a moving
robot %
for a well-founded
ordering. We then conclude using the assumption that the demon is
fair. %

To define the measure, we associate a weight to each of the protocol's phases (see
Section~\ref{sec:unformal_algo})  as follows: 
$  \textrm{Maj} \mapsto 0, \textrm{Dc} \mapsto 1, \textrm{Dd} \mapsto
2, \textrm{Ec, Ic, Sc} \mapsto 3, \textrm{Ed, Id, Sd} \mapsto 4,
\textrm{Gc} \mapsto 5, \textrm{Gd} \mapsto 6.$ Note that
  these weights decrease along the arcs of
  Figure~\ref{fig:graphe}.
 We may now map any
 configuration $C_i$ to a $(p_i,m^{p_i}_i) \in \setN \times \setN$ such
 that $p_i$ is the weight of the phase for the moving robots, and:
 \begin{itemize}
 \item $m^0_i$ is the number of robots that are \emph{not at} the
   unique location of maximal multiplicity, and
 \item $m^{p_i>0}_i$ is 
   $\left\{\begin{array}{ll}  
    \textrm{-} & \textrm{the number of robots that are \emph{not at} target
     if $C_i$ is clean, or}\\
   \textrm{-} & \textrm{the number of robots that are \emph{neither at} target
     \emph{nor} on \SEC if $C_i$ is dirty.}
   \end{array}\right.$
 \end{itemize}

Let $>_\setN$ be the usual ordering on natural numbers, the relevant
ordering $\succ$ is defined as the lexicographic extension of
$>_\setN$ on pairs: 
$
(p,m) \succ (p',m') \textrm{ iff } 
\textrm{either~} p >_\setN p', \textrm{ or }  p =_\setN p' \textrm{ and } m >_\setN m'.
$\\
It is well-founded since $>_\setN$ is well-founded.
We show that for %
any round on a configuration $C_k$ resulting in
a \emph{different} configuration $C_{k+1}$ (that is, some robots have moved), $(p_k,m^{p_k}_k) \succ
(p_{k+1},m^{p_{k+1}}_{k+1})$, hence proving that eventually there is
no more change in successive configurations.

As convincing that they may seem, the arguments
above do not constitute a formal proof at all, and should not be
ultimately relied upon. At best, they may give an intuition that the
protocol is correct. To obtain formal guarantees, we define
the protocol and do the proof in our \coq framework.

\subsection{Formalising the Protocol, Key Points, and the Main Theorem}

\paragraph{Formal description of the protocol.}
\label{sec:concrete-robogram}
The type of locations is $\setR^2$ (noted \lstinline!R2.t! and defined as \lstinline!R*R! from the type \lstinline!R! of the \coq library on axiomatic reals).
The robogram as described in Section~\ref{sec:unformal_algo} is:

\begin{lstlisting}
Definition gatherR2_pgm (s : Spect.t) : R2.t :=
 match Spect.support (Spect.max s) with (* Which are the max height towers?       *)
 | nil => (0, 0) (*@\hfill@*)(* None? only happens when no robot          *)
 | pt :: nil => pt (*@\hfill@*)(* Unique highest tower? go to this tower    *)
 | _ :: _ :: _ => (*@\hfill@*)(* ... otherwise...                         *)
   if is_clean s then target s else (*@\hfill@*)(* If all on SEC/target then go to target *)
   if (0, 0)\in(SECT s) then (0, 0) else target s(*@\hfill@*)(* else cleaning: if on SEC/target*)
 end. (*@\hfill@*)(*  then stay, else go to target   *)
\end{lstlisting}

Target is defined as follows, in critical situations where exactly
three inhabited positions are on the SEC target depends on the shape
of the triangle (here isosceles \emph{excludes} equilateral):
\begin{lstlisting}
Function target_triangle (pt1 pt2 pt3 : R2.t) : R2.t :=
 match classify_triangle pt1 pt2 pt3 with (*@\hfill@*)(* What kind of triangle?   *)
 | Equilateral => barycenter_3_pts pt1 pt2 pt3 (*@\hfill@*)(* To the barycenter        *)
 | Isosceles p => p (*@\hfill@*)(* To the vertex shared by the sides of same lengths *)
 | Scalene => opposite_of_max_side pt1 pt2 pt3 (*@\hfill@*)(* To the vertex that is NOT *)
 end. (*@\hfill@*)(* on the longest side *)
Function target (s : Spect.t) : R2.t :=
 match on_SEC (Spect.support s ) with (*@\hfill@*)(*How many inhabited locations on SEC?*)
 | nil => (0, 0) (*@\hfill@*)(* None? Only happens when no robot   *)
 | pt :: nil => pt (*@\hfill@*)(* Unique loc. on SEC? => gathered!    *)
 | pt1 :: pt2 :: pt3 :: nil => target_triangle pt1 pt2 pt3 (*@\hfill@*)(* See above *)
 | _ => center (SEC l) (*@\hfill@*)(* Gen. case: center of SEC    *)
 end.
\end{lstlisting}
Note that this is almost exactly an actual robot code.
The instantiated robogram (in the sense of Section~\ref{sec:robogram}) binding together this code and its compatibility property is defined under the name \lstinline!gatherR2!.

\paragraph{Formal proofs of key points and of the main theorem.}

The key steps of our proof can be written as relatively straightforward statements.
Theorem \lstinline!round_simplify! expresses the configuration after one round in the global fame of reference, without making reference to local frames of each robot.
Its proof uses several lemmas expressing the invariance of the geometric properties used by the robogram.

Theorem \lstinline!round_simplify! states that to express the
configuration after one round it is correct to use the global demon
view. The global demon view is the spectrum taken directly from the
demon configuration instead of the robot configuration (as defined in
section~\ref{sec:spect}). The statement merges \lstinline!round! and
\lstinline!gatherR2! and forgets about byzantine robots (not
considered in this setting) into a simplified statement:
\begin{lstlisting}
Theorem round_simplify : forall da conf,
  Config.eq (round gatherR2 da conf)
    (fun id =>
       match da.(step) id with
       | None => conf id (*@\hfill@*)(* Robot not activated this round *)
       | Some f =>
           let s := Spect.from_config conf in (*@\hfill@*)(* Take spectrum of demon config *)
           match Spect.support (Spect.max s) with
           | nil => conf id (*@\hfill@*)(* Absurd case: no robots *)
           | pt :: nil => pt (*@\hfill@*)(* Majority stack *)
           | _ => if is_clean s then target s else
                 if mem (conf id) (SECT s) then conf id else target s
           end
       end).
\end{lstlisting}
Where \lstinline!SECT s! corresponds to the union of the \SEC and the
\target of spectrum \lstinline!s!. The proof of this theorem uses the
fact that the robogram uses only operations that are invariant by
change of frame of reference (here applying \lstinline!sim : Sim.t!). For instance the \lstinline!target!
function used in \lstinline!gatherR2! is such:
\begin{lstlisting}
Lemma target_morph : forall (sim : Sim.t) s, Spect.support s <> nil
          -> target (Spect.map sim s) = sim (target s).
\end{lstlisting}

Theorem \lstinline!same_destination! states that two moving robots $id_1$ and
$id_2$ (i.e., that change locations during the round) compute the same destination location (in the demon's frame of reference).
By case on the phases of the robogram, and on the structure of the provided code. The formal proof is about 20 lines 
long and uses Theorem \lstinline!round_simplify!.
\begin{lstlisting}
Theorem same_destination : forall da cf id1 id2,
  In id1 (moving gatherR2 da cf) -> In id2 (moving gatherR2 da cf) 
  -> round gatherR2 da cf id1 = round gatherR2 da cf id2.
\end{lstlisting}

Theorem \lstinline!never_forbidden! says that for all demonic action
\emph{da} and configuration \emph{cf}, if \emph{cf} is not bivalent (\emph{i.e.} not \lstinline!forbidden!),
then the configuration
after the round
is not bivalent.
\begin{lstlisting}
Theorem never_forbidden: forall da cf, 
  \notforbidden cf -> \notforbidden (round gatherR2 da cf).
\end{lstlisting}
The proof is done by a case analysis on the set of towers of maximum
height at the beginning. %
If there is none, this is absurd; %
if there is exactly one, the resulting configuration will have the
same highest tower, a legal configuration. %
Now if there are at least two highest towers, then if the resulting
configuration is bivalent, at least one robot has moved (otherwise the
original configuration would be bivalent, to the contrary of what is
assumed), and all robots that move go to the same of the resulting two
towers. The rest is arithmetics, as described on
page~\pageref{key:not_biv}. The proof of this key point is around 100
lines of \coq script.
Note that as remarked on page~\pageref{key:not_biv} the argument is
the same in $\setR$ or $\setR^2$, hence we were able to reuse the \coq
script developed earlier for~\cite{courtieu15corr} (and thus in our libraries) to prove this
statement. 

It remains to state that for all demonic action \emph{da} and configuration
\emph{conf}, if \emph{conf} is not bivalent, and if there is at least
one robot moving this round, then the configuration resulting from the round
defined by \emph{da} and our robogram on \emph{conf} is smaller than
\emph{conf}. The ordering relation on configurations, called
\lstinline!lt_config!, is the one described in
Section~\ref{key:termination}.

The formal specifications of Gathering have been provided in details
in \cite{courtieu15ipl,courtieu15ipl}. Briefly :
\begin{lstlisting}
(** [gathered_at conf pt] means that in configuration [conf] all good robots
    are at the same location [pt] (exactly). *)
Definition gathered_at (pt: R2.t) (conf: Config.t) := forall g, conf (Good g) = pt.

(** [Gather pt e] means that at all rounds of (infinite) execution
    [e], robots are gathered at the same position [pt]. *)
CoInductive gather (pt: R2.t) (e : execution) : Prop :=
  Gathering : gathered_at pt (execution_head e)
              -> gather pt (execution_tail e)
              -> gather pt e.

(** [WillGather pt e] means that infinite execution [e] is eventually Gathered. *)
Inductive willGather (pt : R2.t) (e : execution) : Prop :=
  | Now : gather pt e -> willGather pt e
  | Later : willGather pt (execution_tail e) -> willGather pt e.
\end{lstlisting}

The theorem stating the correctness of our robogram is then simply:
for all demon $d$ that is fair,
\lstinline!gatherR2! is a solution with reference to $d$.
\begin{lstlisting}
Theorem round_lt_config: forall da conf, \not forbidden conf
    -> moving gatherR2 da conf <> nil -> lt_config (round gatherR2 da conf) conf.
Theorem Gathering_in_R2 : forall d, Fair d -> solGathering gatherR2 d.
\end{lstlisting} 

The proof is led by well-founded induction on the
\lstinline!lt_config!  relation. If all robots are gathered, then it
is done. If not, by fairness some robots will have to move, thus a
robot will be amongst the first to move. (Formally, this is an
induction using fairness.)
We conclude by using the induction hypothesis (of our well-founded
induction) as this round decreases the measure on configurations
(theorem \lstinline!round_lt_config!).
This proof of the main theorem is interestingly small as it is only
20 lines long.
The whole file dedicated to specification and certification of our
algorithm (\lstinline!Algorithm.v!) consists of
478~lines of definitions, specification and intermediate lemmas, and
2836~lines of actual proof.

\paragraph{Axioms of the formalisation}

At the end of the main file \lstinline!Algorithm.v! can be found a printing command,
 \lstinline!Print Assumptions Gathering_in_R2!, showing all the axioms upon which the proof of correctness of our algorithm for gathering in $\setR^2$ relies, in total 39 axioms.
Here, we break them down.
They can be classified in four categories:
\begin{itemize}
\item The first category is the axiomatisation of reals numbers from the \coq standard library. It is by far the biggest number of axioms (26), and they are not listed here.
\item The second category is the description of the problem.
  \begin{lstlisting}
    nG : nat
    nG_conf : 3 <= nG
  \end{lstlisting}
  As one can see, it simply means that our proof is valid for any number~\lstinline!nG! of robots greater than 2. Notice that with 2 or less robots, the problem is not very interesting:
  \begin{itemize}
  \item With one robot or less, the problem is not interesting.
  \item With 2 robots, either they start gathered and there is nothing to do, or they are in different locations, which is a forbidden configuration for which an impossibility results exists \cite{suzuki99siam}.
  \end{itemize}
\item The third category is the axiomatisation of the SEC: we do not give an algorithm computing it but instead axiomatize it.
  \begin{lstlisting}
    SEC : list R2.t -> circle
    SEC_compat : Proper (Permutation (A:=R2.t) ==> eq) SEC
    SEC_spec1 : forall l : list R2.t, enclosing_circle (SEC l) l
    SEC_spec2 : forall (l : list R2.t) (c : circle),
                enclosing_circle c l -> (radius (SEC l) <= radius c)%R
    SEC_nil : radius (SEC nil) = 0%R
  \end{lstlisting}
  Axiom \lstinline!SEC_compat! expresses that the SEC does not depend on the order in which points are given.
  The other three axioms are the specification of the SEC: it is an enclosing circle (\lstinline!SEC_spec1!) and the smallest one (\lstinline!SEC_spec2!).
  The last one (\lstinline!SEC_nil!) is used to fill the corner case where the other two axioms are not enough. \\
  There are also three geometric properties of SEC that we think could be provable from its axiomatisation but are currently left as axioms:
  \begin{lstlisting}
    SEC_unicity : forall (l : list R2.t) (c : circle),
                  enclosing_circle c l ->
                  (radius c <= radius (SEC l))%R -> c = SEC l
    equilateral_SEC : forall pt1 pt2 pt3 : R2.t,
                      classify_triangle pt1 pt2 pt3 = Equilateral ->
                      SEC (pt1 :: pt2 :: pt3 :: nil) =
                      {| center := barycenter_3_pts pt1 pt2 pt3;
                         radius := R2.dist (barycenter_3_pts pt1 pt2 pt3) pt1 |}
    SEC_on_SEC : forall l,  SEC l = SEC (on_SEC l) .
  \end{lstlisting}
  The first one expresses that a SEC is unique (note that this property is not true with all distances, but it is with the euclidean one); the second one gives the SEC of an equilateral triangle; the last one says that the points strictly inside the SEC can be removed without changing the SEC.
\item The fourth category are usual geometric properties that are not part of our library.
  We expect to be able to use another geometric library (for instance via Euclid axiomatisation) to get some proofs.
  \begin{lstlisting}
    bary3_unique : forall x y z a b : R2.t,
                   is_barycenter_3_pts x y z a ->
                   is_barycenter_3_pts x y z b -> R2.eq a b
    bary3_spec : forall pt1 pt2 pt3 : R2.t,
                 is_barycenter_3_pts pt1 pt2 pt3 (barycenter_3_pts pt1 pt2 pt3)
    three_points_same_circle : forall (c1 c2 : circle) (pt1 pt2 pt3 : R2.t),
                               NoDup (pt1 :: pt2 :: pt3 :: nil) ->
                               on_circle c1 pt1 = true ->
                               on_circle c1 pt2 = true ->
                               on_circle c1 pt3 = true ->
                               on_circle c2 pt1 = true ->
                               on_circle c2 pt2 = true ->
                               on_circle c2 pt3 = true -> c1 = c2
  \end{lstlisting}
  The first two link two possible definitions of the barycenter of three points:
  \begin{itemize}
  \item the mean of the points (definition \lstinline!barycenter_3_pts!),
  \item the point that minimises the sum of the square of the distances to these three points (property \lstinline!is_barycenter_3_pts!).
  \end{itemize}
  The first axiom states that the second definition defines a unique point and the second one that both definitions coincide. \\
  The last axiom states that through three points, at most one circle can run.
  Note that there can be none if the points are aligned.
\end{itemize}

\section{Discussion and Perspectives}\label{sec:discussion}

The Distributed Computing community is known to have fundamental algorithms tightly coupled with their proof of correctness. The mobile robot setting is no exception, as the minimal hypotheses a protocol must make to solve a given problem are extremely difficult to identify without actually writing the corresponding correctness proofs (that is, an intuitive approach is often detrimental to the correctness of the result to be established, as recent errors found in the literature proved~\cite{adamek15edcc}). In a formal proof approach to obtain mechanically certified protocols, our framework and methodology clearly contributes to two main phases in a verified development. %

Firstly the \emph{specification} phase, where all objects, definitions,
algorithms, statements and expected properties are expressed without
any ambiguity, in a higher order type theoretic functional
environment. The lack of ambiguity is a key feature to enable the
early detection of inconsistencies between the problem specification,
the algorithmic proposal, and the execution model. We emphasise the
fact that there is no need to be an expert with the \coq proof assistant to use our framework in this phase.
Clear and unequivocal specifications are indeed a fundamental step towards correct algorithms. \todoLR{citer Lamport}

Secondly the \emph{proof} phase, where properties are proved to hold for the relevant executions. This phase is of course more demanding on the expertise side, so our goal when constructing the framework was to provide useful libraries and proof techniques that can be reused in other contexts, enabling more automation to the protocol designer. Considering reusability, useful assets brought by the current work are the notions of gathering, SSYNC demons, etc., developments on geometry in $\setR^2$ and smallest enclosing circles, and the proof of \lstinline!never_forbidden!~\cite{courtieu15corr}. Those will most likely prove useful in future developments. When developing the protocol for our case study, we decided to modify the protocol code several times, either to fix a newly discovered bug, or to ease the writeup of the proofs. This classical design stage was streamlined by the use of a formal language based on the Curry-Howard isomorphism \cite{howard80ecllcf} where both activities can be done in a uniform way. In such a setting, correcting the algorithm amounts to modifying the algorithm definition, and replaying the proofs certification process after adapting the proof scripts written previously. The mechanised verification of the proofs makes this process fast and trustworthy, compared to a purely handcrafted approach.

\paragraph{Perspectives}\label{sec:conclu}

A next step would be to add more dimensions to the considered
Euclidean space. %
As the framework is highly parametric, specifying another
space in which robots move is not a dramatic change: the type of
locations is a parameter, it is left abstract throughout the majority
of the formalism, in which a concrete instance is not
needed. %
Another interesting evolution would be to take into account the more
general ASYNC model, that is when Look-Compute-Move cycles and stages are not atomic anymore. %
Describing behaviours that are ASYNC in \coq may nonetheless
add to the intricacy of formal proofs, and relevant libraries to ease
the task of the developer will have to be provided accordingly. 





\begin{thebibliography}{10}

\bibitem{adamek15edcc}
Jordan Adamek, Mikhail Nesterenko, and S{\'{e}}bastien Tixeuil.
\newblock Evaluating and optimizing stabilizing dining philosophers.
\newblock In {\em 11th European Dependable Computing Conference, {EDCC} 2015,
  Paris, France, September 7-11, 2015}, pages 233--244. {IEEE}, 2015.

\bibitem{auger13sss}
C\'edric Auger, Zohir Bouzid, Pierre Courtieu, S\'ebastien Tixeuil, and Xavier
  Urbain.
\newblock {Certified Impossibility Results for Byzantine-Tolerant Mobile
  Robots}.
\newblock In Teruo Higashino, Yoshiaki Katayama, Toshimitsu Masuzawa, Maria
  Potop-Butucaru, and Masafumi Yamashita, editors, {\em Stabilization, Safety,
  and Security of Distributed Systems - 15th International Symposium (SSS
  2013)}, volume 8255 of {\em Lecture Notes in Computer Science}, pages
  178--186, Osaka, Japan, November 2013. Springer-Verlag.

\bibitem{BMPTT13r}
B{\'e}atrice Berard, Laure Millet, Maria Potop-Butucaru, Yann Thierry-Mieg, and
  S{\'e}bastien Tixeuil.
\newblock {Formal verification of Mobile Robot Protocols}.
\newblock Technical report, May 2013.

\bibitem{bertot04coqart}
Yves Bertot and Pierre Cast\'eran.
\newblock {\em Interactive Theorem Proving and Program Development. Coq'Art:
  The Calculus of Inductive Constructions}.
\newblock Texts in Theoretical Computer Science. Springer-Verlag, 2004.

\bibitem{bonnet14wssr}
Fran{\c{c}}ois Bonnet, Xavier D{\'{e}}fago, Franck Petit, Maria
  Potop{-}Butucaru, and S{\'{e}}bastien Tixeuil.
\newblock Discovering and assessing fine-grained metrics in robot networks
  protocols.
\newblock In {\em 33rd {IEEE} International Symposium on Reliable Distributed
  Systems Workshops, {SRDS} Workshops 2014, Nara, Japan, October 6-9, 2014},
  pages 50--59. {IEEE}, 2014.

\bibitem{bouzid13icdcs}
Zohir Bouzid, Shantanu Das, and S{\'{e}}bastien Tixeuil.
\newblock Gathering of mobile robots tolerating multiple crash faults.
\newblock In {\em ICDCS}, pages 337--346, Philadelphia, Pennsylvania, USA, July
  2013. IEEE Computer Society.

\bibitem{bramas15sirocco}
Quentin Bramas and S{\'{e}}bastien Tixeuil.
\newblock Wait-free gathering without chirality.
\newblock In Christian Scheideler, editor, {\em Structural Information and
  Communication Complexity - 22nd International Colloquium, {SIROCCO} 2015,
  Montserrat, Spain, July 14-16, 2015, Post-Proceedings}, volume 9439 of {\em
  Lecture Notes in Computer Science}, pages 313--327. Springer, 2015.

\bibitem{berard15infsoc}
Béatrice Bérard, Pierre Courtieu, Laure Millet, Maria Potop-Butucaru, Lionel
  Rieg, Nathalie Sznajder, Sébastien Tixeuil, and Xavier Urbain.
\newblock {Formal Methods for Mobile Robots: Current Results and Open
  Problems}.
\newblock {\em International Journal of Informatics Society}, 7(3):101--114,
  2015.
\newblock Invited Paper.

\bibitem{coquand90colog}
Thierry Coquand and Christine Paulin-Mohring.
\newblock {Inductively Defined Types}.
\newblock In Per Martin-L{\"o}f and Grigori Mints, editors, {\em {International
  Conference on Computer Logic ({C}olog'88)}}, volume 417 of {\em Lecture Notes
  in Computer Science}, pages 50--66. Springer-Verlag, 1990.

\bibitem{courtieu15corr}
Pierre Courtieu, Lionel Rieg, Sébastien Tixeuil, and Xavier Urbain.
\newblock {A Certified Universal Gathering Algorithm for Oblivious Mobile
  Robots}.
\newblock {\em CoRR}, abs/1506.01603, 2015.

\bibitem{courtieu15ipl}
Pierre Courtieu, Lionel Rieg, Sébastien Tixeuil, and Xavier Urbain.
\newblock {Impossibility of Gathering, a Certification}.
\newblock {\em Information Processing Letters}, 115:447--452, 2015.

\bibitem{devismes12sss}
St\'{e}phane Devismes, Anissa Lamani, Franck Petit, Pascal Raymond, and
  S\'{e}bastien Tixeuil.
\newblock {Optimal Grid Exploration by Asynchronous Oblivious Robots}.
\newblock In Andr{\'e}a~W. Richa and Christian Scheideler, editors, {\em
  Stabilization, Safety, and Security of Distributed Systems - 14th
  International Symposium (SSS 2012)}, volume 7596 of {\em Lecture Notes in
  Computer Science}, pages 64--76, Toronto, Canada, October 2012.
  Springer-Verlag.

\bibitem{dieudonne12tcs}
Yoann Dieudonn{\'{e}} and Franck Petit.
\newblock Self-stabilizing gathering with strong multiplicity detection.
\newblock {\em Theoretical Computer Science}, 428:47--57, 2012.

\bibitem{flocchini12book}
Paola Flocchini, Giuseppe Prencipe, and Nicola Santoro.
\newblock {\em {Distributed Computing by Oblivious Mobile Robots}}.
\newblock Synthesis Lectures on Distributed Computing Theory. Morgan {\&}
  Claypool Publishers, 2012.

\bibitem{howard80ecllcf}
William~A. Howard.
\newblock The formulae-as-types notion of construction.
\newblock In J.~Roger~Hindley Jonathan P.~Seldin, editor, {\em To H. B. Curry:
  Essays on Combinatory Logic, Lambda Calculus and Formalism}, pages 479--490.
  Academic Press, London, 1980.

\bibitem{izumi13tpds}
Taisuke Izumi, Tomoko Izumi, Sayaka Kamei, and Fukuhito Ooshita.
\newblock Feasibility of polynomial-time randomized gathering for oblivious
  mobile robots.
\newblock {\em IEEE Transactions on Parallel and Distributed Systems},
  24(4):716--723, 2013.

\bibitem{izumi12siam}
Taisuke Izumi, Samia Souissi, Yoshiaki Katayama, Nobuhiro Inuzuka, Xavier
  D{\'{e}}fago, Koichi Wada, and Masafumi Yamashita.
\newblock The gathering problem for two oblivious robots with unreliable
  compasses.
\newblock {\em SIAM Journal of Computing}, 41(1):26--46, 2012.

\bibitem{MPST14c}
Laure Millet, Maria Potop{-}Butucaru, Nathalie Sznajder, and S{\'{e}}bastien
  Tixeuil.
\newblock On the synthesis of mobile robots algorithms: The case of ring
  gathering.
\newblock In Pascal Felber and Vijay~K. Garg, editors, {\em Stabilization,
  Safety, and Security of Distributed Systems - 16th International Symposium,
  (SSS 2014)}, volume 8756 of {\em Lecture Notes in Computer Science}, pages
  237--251, Paderborn, Germany, sep 2014. Springer-Verlag.

\bibitem{sangiorgi12book}
Davide Sangiorgi.
\newblock {\em {Introduction to Bisimulation and Coinduction}}.
\newblock Cambridge University Press, 2012.

\bibitem{suzuki99siam}
Ichiro Suzuki and Masafumi Yamashita.
\newblock {Distributed Anonymous Mobile Robots: Formation of Geometric
  Patterns}.
\newblock {\em SIAM Journal of Computing}, 28(4):1347--1363, 1999.

\end{thebibliography}
\end{document}